# RANDOM-SET METHODS IDENTIFY DISTINCT ASPECTS OF THE ENRICHMENT SIGNAL IN GENE-SET ANALYSIS

BY MICHAEL A. NEWTON, FERNANDO A. QUINTANA, JOHAN A. DEN BOON, SRIKUMAR SENGUPTA AND PAUL AHLQUIST

*University of Wisconsin–Madison, Pontificia Universidad Católica de Chile, University of Wisconsin–Madison, WiCell Research Institute, and University of Wisconsin–Madison and Howard Hughes Medical Institute*

A prespecified set of genes may be enriched, to varying degrees, for genes that have altered expression levels relative to two or more states of a cell. Knowing the enrichment of gene sets defined by functional categories, such as gene ontology (GO) annotations, is valuable for analyzing the biological signals in microarray expression data. A common approach to measuring enrichment is by cross-classifying genes according to membership in a functional category and membership on a selected list of significantly altered genes. A small Fisher's exact test $p$-value, for example, in this $2 \times 2$ table is indicative of enrichment. Other category analysis methods retain the quantitative gene-level scores and measure significance by referring a category-level statistic to a permutation distribution associated with the original differential expression problem. We describe a class of random-set scoring methods that measure distinct components of the enrichment signal. The class includes Fisher's test based on selected genes and also tests that average gene-level evidence across the category. Averaging and selection methods are compared empirically using Affymetrix data on expression in nasopharyngeal cancer tissue, and theoretically using a location model of differential expression. We find that each method has a domain of superiority in the state space of enrichment problems, and that both methods have benefits in practice. Our analysis also addresses two problems related to multiple-category inference, namely, that equally enriched categories are not detected with equal probability if they are of different sizes,

Received October 2006; revised February 2007.
[1]Supported by the National Cancer Institute Grants CA64364, CA22443 and CA97944 and FONDECYT Grants 1020712 and 1060729.

Supplementary material available at http://imstat.org/aoas/supplements

*Key words and phrases.* Conditional testing, gene ontology, gene set enrichment analysis, host-virus association in nasopharyngeal carcinoma, selection versus average evidence, significance analysis of function and expression.

and also that there is dependence among category statistics owing
to shared genes. Random-set enrichment calculations do not require
Monte Carlo for implementation. They are made available in the R
package *allez*.

**1. Introduction.** In processing results of a microarray study, one is faced
with the daunting task of relating differential-expression evidence to other
information about the genes. Any interesting connections that can be revealed are critical in developing a fuller understanding of the biology and in
providing guidance toward the next experiment [e.g., Rhodes and Chinnaiyan
(2005)]. Much of the exogenous information is organized in networks of functional categories; genes are annotated to the same category by virtue of a
shared biological property. The Gene Ontology (GO) project is perhaps the
best example of how biological information is carried by networked collections of functional categories [Gene Ontology Consortium (2000, 2004)]. Initiated as a collaboration among different genome projects, GO has become
a fundamental resource that records attributes of genes and gene products
and that organizes these attributes using networks of connected functional
categories.

The problem of enrichment emerges in relating gene-level expression results with functional categories. To what extent, if at all, are genes with
altered expression over-represented in a named category? At the risk of oversimplifying things, the extensive research and development toward solving
this problem may be classified by two statistical approaches. The first begins
by selecting a short list of genes that are altered significantly relative to the
cell grouping under study: for instance, genes with extreme fold change or
with extreme value of a test statistic. The intersection of the selected list
and the functional category is then evaluated, perhaps by Fisher's exact test
or a variant, which scores the category highly for enrichment if many more
selected genes than expected belong to the category [Drăghici et al. (2003),
Berriz et al. (2003), Doniger et al. (2003), Al-Shahrour, Uriarte and Dopazo
(2004), Beißbarth and Speed (2004), Cheng et al. (2004), Zhong et al. (2004),
Dodd et al. (2006)]. Available informatics tools and related problems are reviewed in Khatri and Drăghici (2005). A second approach is developed in
Virtaneva et al. (2001) and Barry, Nobel and Wright (2005), called SAFE
(Significance analysis of function and expression) and also in Mootha et al.
(2003) and Subramanian et al. (2005), called GSEA (Gene set enrichment
analysis). Briefly, expression information on all the genes under study is retained; then a permutation analysis is used to measure the significance of
category-level statistics computed from these gene-level statistics.

Existing tools have been effective in adding value to expression results,
but they remain limited for evaluating enrichment signals. Analysis is simplified when considering selected gene lists, since quantitative scores from the gene-level analysis are not required. But then the enrichment results depend on the stringency of the selection, and give equal weight to genes at both ends of the selected list. This problem is redressed in the SAFE/GSEA approach. The permutation method adopted by SAFE/GSEA refers back to the labeled microarray data themselves rather than to the results of the differential-expression analysis. There is an added computational burden in this strategy and also it can become ineffective when few microarrays enter the permutation. A technical issue, further, concerns the null hypothesis at work in the SAFE/GSEA permutation. It refers to the complete absence of differential expression rather than to the absence of enrichment.

In this paper we explore properties of *random-set* methods for measuring enrichment. We adopt category-level statistics like in SAFE/GSEA, but we calibrate them in the same way that Fisher's exact test calibrates the intersection of a functional category and a selected list. That is, we calibrate them conditionally on results of the differential expression analysis by considering values of the category-level statistic that would be achieved by a random set of genes (Section 2). Calculations are simplified by formulae for the expected value and variance of this conditional distribution, so that Monte Carlo approximations may not be required. Random-set scoring is applicable to a variety of gene-level scores; we compare two schemes empirically in a study of nasopharyngeal cancer in Section 3. One measures enrichment by counting the intersection with a selected gene list; the other considers average differential-expression evidence across all genes in the category. In conjunction with empirical evidence we pursue a theoretical analysis to compare these two category scoring methods (Section 4). We find that two parameters affect the power to detect enrichment, and these play out so that neither the selection approach nor the averaging approach is uniformly superior. Additionally, we show how the random-set approach facilitates simultaneous inference among multiple categories. Two important issues are (1) how to accommodate the power imbalance caused by differently sized-categories, and (2) how to obtain the joint distribution of category scores in order to have valid type-I error rate control (Section 5). We offer approximate analytical solutions to these problems.

**2. Random-set enrichment scoring.** We describe a general method to score categories for enrichment with expression-altered genes. Sengupta et al. (2006) (especially Supplementary Data) introduced the method and described it briefly. It forms the basis of our approach and so here we amplify and clarify the presentation. The class extends Fisher's exact test by allowing a variety of gene-level scores, denoted $\{s_g\}$, for different genes $g$. These may be binary indicators of extreme differential expression, but we allow more general quantitative expression scores. We focus initially on a single category $C$ containing $m$ genes.

The idea is to consider the unstandardized enrichment score $\bar{X} = \frac{1}{m}\sum_{g \in C} s_g$ as a random variable wherein the randomness comes not through the gene scores $\{s_g\}$ but rather through the set $C$. We are concerned, after all, with measuring enrichment for a specific category $C$ compared to other hypothetical categories from the same system. It is useful to treat the random set $C$ as drawn uniformly at random from the $\binom{G}{m}$ subsets of $m$ distinct genes from the population of $G$ genes, just a simple random sample drawn without replacement [e.g., Cochran (1977)]. This is equivalent to a permutation scheme in which gene-level scores are randomly shuffled among the gene labels. Precisely this scheme underlies Fisher's exact test in the special case that $s_g$ is the binary indicator of selection onto the significantly altered gene list (Supplementary Table 1). The random-set model is applicable beyond the binary case to any sort of gene-level scores, though the induced distribution of $\bar{X}$ becomes intractable. Rather than resort to Monte Carlo, we find that the first two moments of the otherwise intractable distribution are available analytically, and that the induced distribution is approximately Gaussian. These findings are the basis of our proposed standardization. Under the random-set model, and thus conditional on gene-level scores $\{s_g\}$,

$$\mu = E(\bar{X}) = \frac{\sum_{g=1}^G s_g}{G} \tag{1}$$

and

$$\sigma^2 = \mathrm{var}(\bar{X}) = \frac{1}{m}\left(\frac{G-m}{G-1}\right)\left\{\left(\frac{\sum_{g=1}^G s_g^2}{G}\right) - \left(\frac{\sum_{g=1}^G s_g}{G}\right)^2\right\}, \tag{2}$$

which are easily computed from the full set of gene-level scores and the category size. Notably, the mean $\mu$ does not depend on attributes of the category, though the variance depends on the category size $m$. We propose the standardized category-enrichment score $Z = (\bar{X} - \mu)/\sigma$, which is a mean zero, unit variance score on the null hypothesis that category $C$ is not enriched for differentially expressed genes. Analysis is simplified, especially in the case of multiple categories, because $Z$ is computable without using permutation. Large values of $Z$ favor the enrichment hypothesis. For moderate to large categories, central limit theory indicates that $Z$ is distributed approximately as a standard normal on the no-enrichment null hypothesis.

Enrichment scoring is enlivened by the possibility of using a variety of gene-level scores. We may use log fold changes, $t$-statistics or other local measures of differential expression. In the special case where $\{s_g\}$ are the ranks associated with gene-level scores, we get a version of the Wilcoxon test for enrichment, since $m\bar{X}$ is a sum of ranks, and both $\mu$ and $\sigma^2$ simplify as

$$\mu = \frac{G+1}{2}, \qquad \sigma^2 = \frac{(G-m)(G+1)}{12m},$$

with suitable adjustments for ties. Gene-level scores from an empirical Bayesian analysis might be posterior probabilities of differential expression [Kendziorski et al. (2003)], in which case $m\bar{X}$ equals the posterior expected number of altered genes in the category, and $Z$ calibrates this relative to the population of genes. Section 3 develops an example in which $\{s_g\}$ are transformed Spearman correlations between host genes and the expression of a particular viral gene. Efficiency and approximate normality of the $Z$ score will be improved if the distribution of gene-level scores is suitably regular. For instance, it is preferred to use log transformed $p$-values instead of $p$-values, and log fold instead of raw fold change.

Another important special case happens when $\{s_g\}$ are binary scores indicating selection to a short list of significantly differentially expressed genes. Then $Z^2 = (\frac{G-1}{G})U$, where $U$ is Pearson's chi-squared statistic for testing independence between category and short-list assignment [calculations not shown, but following, e.g., Bickel and Doksum (2001), page 402]. To a minor approximation, then, our proposed $Z$ score corresponds to Fisher's or Pearson's test when $\{s_g\}$ are binary gene-level scores. Otherwise it generalizes those category scores and measures other aspects of the enrichment signal, as we demonstrate next.

**3. An analysis of host/virus associations in cancer.** A recent expression study of nasopharyngeal carcinoma (NPC) used the proposed methodology for category enrichment [Sengupta et al. (2006)]. NPC is a cancer of the nasopharynx that is responsible for 60–70,000 deaths per year worldwide. Nearly all cases are associated with the Epstein–Barr virus (EBV) infection, though the molecular determinants and the nature of the host-virus interactions remain poorly understood. Sengupta et al. (2006) studied tumor tissue from $n = 31$ NPC patients using Affymetrix hgu133plus2 microarrays to measure host gene expression and using RT-PCR to measure the expression of 10 viral genes. The hgu133plus2 microarrays probe the transcriptome with $G = 54,675$ probe sets. Here we reconsider associations between host expression and the expression of the single viral gene EBNA1.

The statistical analysis of host-virus association rests on pairwise Spearman correlations between individual Affymetrix probe set values and the expression of EBNA1. Supplementary Figure 1 shows one probe set and its correlation with EBNA1. Supplementary Figure 2 shows correlations with EBNA1 for all host probe sets. The most extreme negative correlation is $r = -0.75$, which is unusually small (Supplementary Figure 3, $p$-value $= 0.04$). A striking feature of the empirical distribution of correlations is that 65% of host probe sets are negatively correlated with EBNA1. This is significantly more than expected if truly there is no association between host and virus expression ($p$-value $= 6 \times 10^{-4}$, Supplementary Figure 3). Globally, there is evidence for significant negative association between EBNA1 expression and the expression of host genes in NPC. Figure 1 highlights a selected list of the 574 most significantly altered host probe sets; the list targets a 5% false discovery rate (FDR) according to the $q$-value method of Storey (2003). In this calculation $p$-values were obtained by recalling that

$$(3) \qquad s_g = \frac{1}{2}\sqrt{n-3}\log\frac{1-r_g}{1+r_g}$$

is approximately standard normal in the absence of a true correlation between $g$ and EBNA1 [Fisher (1921)]. The sign change employed (compared to the usual inverse hyperbolic tangent transform) means that genes which correlate negatively with EBNA1 have a positive gene score $s_g$. Naturally, we may examine the genes on this selected list, but a study of functional categories that are enriched for negatively associated genes exposes more of the relevant biology.

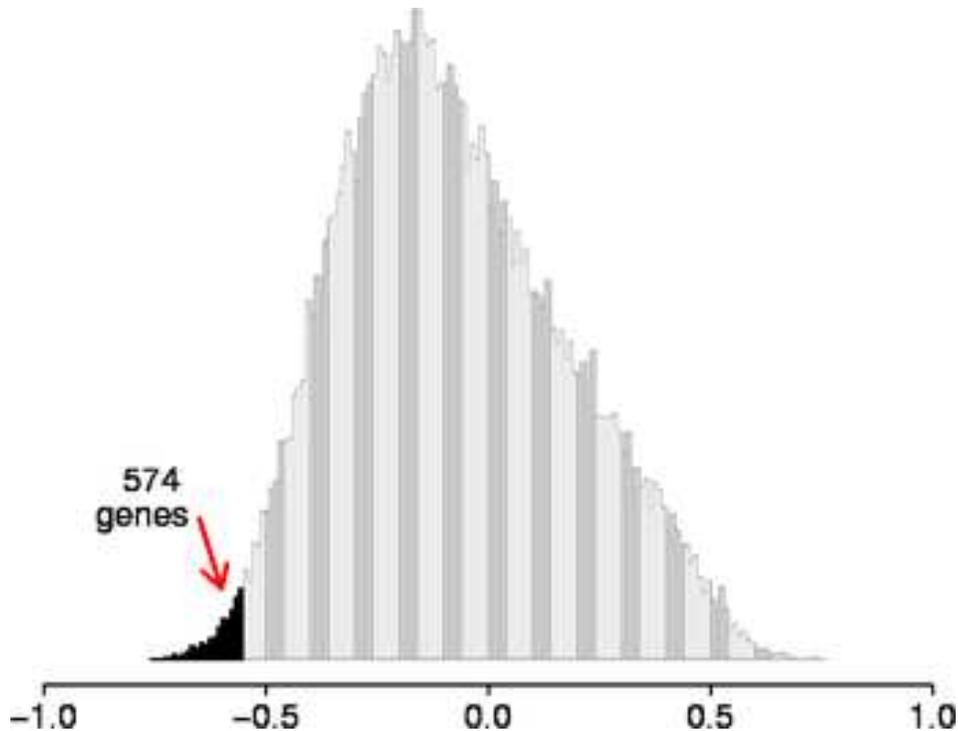

Fig. 1. *Gene selection. For the host-virus association example, plotted is a histogram of the 54,675 Spearman correlations between expression of each Affymetrix probe set and the expression of the viral gene EBNA1. Correlations are computed using microarray data on* 31 *tumor samples as described in Sengupta et al. (2006). Highlighted on the left are selected probe sets that have significant negative correlation with EBNA1 according to a q-value analysis that targets a 5% FDR gene list (correlation less than* −0.55*).*

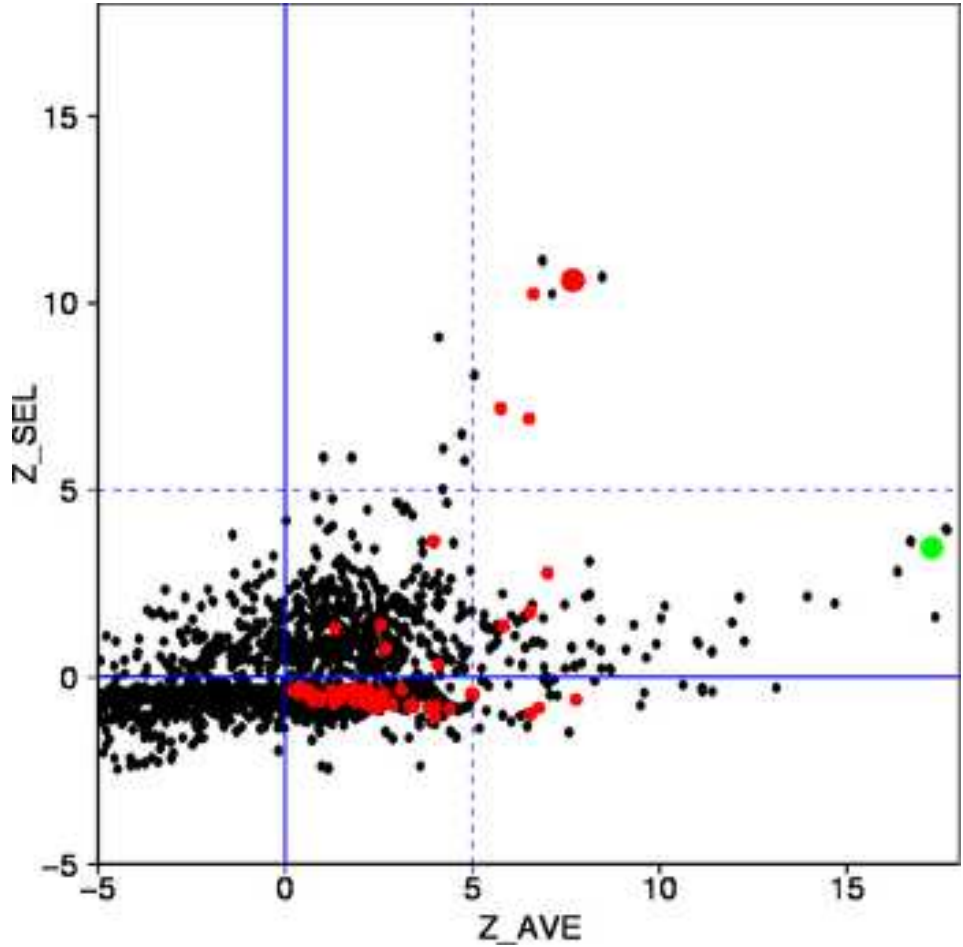

FIG. 2. *Random-set scoring. Shown are results of two category-scoring methods applied to 2761 GO categories for which $m \geq 10$ and based on the gene-level correlations from Figure 1. Both methods aim to detect enrichment of the category for genes that are negatively associated with EBNA1 viral expression. Category GO:0019883, "antigen presentation, endogenous antigen" contains $m = 48$ probe sets and scores highly by both methods (large red dot). The large $(m = 1494)$ "immune response" category (GO:0006955) scores extremely highly by $Z_{\mathrm{ave}}$ but not so highly by $Z_{\mathrm{sel}}$ (green dot). Shown in red are categories that are subsets of "immune response." The average correlation of immune response genes with EBNA1 is extremely significant, but the number of significantly negatively correlated genes is less significant.*

Figure 2 summarizes two category-enrichment scoring methods applied to all GO categories (2761) containing at least $m = 10$ annotated hgu133plus2 probe sets. (This used the October 2005 build of Bioconductor package hgu133plus2.) Many probe sets were unannotated, and to avoid potential biases, we restricted attention to the universe of $G = 27{,}152$ annotated probe sets [Al-Shahrour, Uriarte and Dopazo (2004)]. The two enrichment scoring methods are conditional $Z$ scores as described in Section 2. The first, $Z_{\text{ave}}$, is based on gene-level transformed correlations $s_g$ from (3). The second, $Z_{\text{sel}}$, is based on binary scores $1[s_g > k]$, where $k$ defines the 5% FDR list of the most significantly negatively correlated host genes. Recall that $Z_{\text{sel}}$ is the normal-score version of Fisher's exact test. Since each $Z$ score is nominally standard normal in the absence of enrichment, Figure 2 seems to indicate that many GO categories are enriched for altered genes. Reference lines at $z = 5$ are drawn for guidance (nominal $p$-value $< 10^{-6}$). A noteworthy feature in Figure 2 is that $Z_{\text{sel}}$ and $Z_{\text{ave}}$ are not perfectly correlated. They capture different aspects of the enrichment signal, and thus, they deserve separate consideration. Some categories have high $Z_{\text{sel}}$ but negligible $Z_{\text{ave}}$. They are enriched for genes on the short list of most significantly negatively correlated genes, but the average correlation is not unusual. Other categories have high $Z_{\text{ave}}$ but negligible $Z_{\text{sel}}$. These would not be detected by Fisher's test, for example, though on the average the negative correlation exhibited by the contained genes is extremely unusual.

That $Z_{\text{sel}}$ and $Z_{\text{ave}}$ capture different aspects of the enrichment signal is exemplified by the immune response category, GO:0006955, which connects $m = 1494$ probe sets on the hgu133plus2 microarray. Recall that a significant mass of host probe sets are negatively correlated with the EBNA1, though most of these do not occupy the 5% FDR selected list of most significantly altered host probe sets. Among the 2761 GO categories are many (marked in red) that are subsets of GO:006955; that is, they represent specific forms of the immune response. Notably, all these subsets have $Z_{\text{ave}} > 0$, which indicates that their average correlation with EBNA1 is more negative than average. Taken together, we get strong evidence of enrichment by $Z_{\text{ave}}$. At the same time, many of the subsets have $Z_{\text{sel}} < 0$, which indicates that they have less representation on the selected list than we expect, and thus, selected probe sets are not particularly over-represented in the immune response category.

Sengupta et al. (2006) followed up on some of the categories that showed both extreme $Z_{\text{sel}}$ and extreme $Z_{\text{ave}}$, such as GO:0019883, which is in the biological process network, with GO term *antigen presentation, endogenous antigen*. This category is marked up in subsequent figures. There are $m = 48$ probe sets annotated to this category, and $x = 8$ occupy the selected list, giving $z_{\text{sel}} = 10.6$. Also, the average correlation with EBNA1 is unusually low, with $z_{\text{ave}} = 7.68$. An informal look at the short list of 574 significantly altered probe sets probably would not have revealed a preponderance of GO:0019883 genes. Indeed, the best ranking is at the 90th position, there are only three probe sets in the top 250. Followup experiments on the genes in GO:0019883 confirmed the negative correlation findings that were suggested by the enrichment analysis. Ongoing research aims to understand whether viral EBNA1 is taking advantage of host cells that have disabled antigen presentation function, or whether the virus is effecting a change in the host expression itself.

It is routine that named genes are associated with multiple probe sets on an Affymetrix microarray (Supplementary Figure 4A). GO:0019883, for example, represents only $m_g = 12$ genes though it has $m_p = 48$ probe sets. The fact is important for enrichment calculations since we ought to avoid spurious findings that reflect over-representation of certain genes in the system rather than biologically significant enrichment. Various solutions are available. Ideally we would first reduce the probe set data to the gene level, and then proceed with enrichment calculations on this reduced space (giving $Z_{\text{ideal}}$). A computationally much simpler adjustment is suggested by the variance formula (2). It uses the probe set based $Z$ score, and the numbers $m_g$ and $m_p$ to compute

$$(4) \qquad Z_{\text{adjust}} = Z \sqrt{\frac{m_g}{m_p} \frac{(G - m_p)}{(G - m_g)}}.$$

The rationale is that $\bar{X}$ and $\mu$ may not change much in the reduction step; most of the effect will be on the variance. The naive, ideal (reduce by median) and adjusted enrichment scores are compared in Supplementary Figure 4B, C. The ideal scores tend to be more conservative than the naive ones; GO:0019883 remains impressive with $z_{\text{ideal;ave}} = 4.78$ and $z_{\text{ideal;sel}} = 11.3$ (latter not shown). Globally, the adjustment (4) is similar to the ideal score and it tends to be conservative. In the example category GO:0019883, $z_{\text{adjust;ave}} = 3.84$ and $z_{\text{adjust;sel}} = 5.3$. Thus, effective approximations accommodate the multiple probe sets per gene problem. In some cases it may be possible to select the most reliable probe sets from among the multiple probe sets associated with a gene. For example, a probe set showing high average intensity and high dynamic range across multiple tissue types may be better than one measuring near the background signal in most samples. We do not address the selection of probe sets in this paper, but if probe sets are selected for some genes, equation (4) could be applied to the selected probe sets.

For further comparison, we applied the SAFE procedure [Barry, Nobel and Wright (2005)] to all 2761 GO categories using the original microarray data and EBNA1 expression data on all 31 tumor samples. We adopted the same category-level statistic in order to control the comparison. Specifically, gene-level Spearman correlations were transformed to ranks to be used as $s_g$ values, and the category-level statistic $\bar{X}$ was the average rank (ranks were relative to the 27,152 probe sets having some annotation). Results for GO:0019883 are summarized in Figure 3. Visually, the category appears to have a preponderance of negative correlations with EBNA1 (Panel A), and this is supported by both statistical calibrations. Yet random-sets and SAFE evaluate the significance of the same average-rank statistic rather differently. Compared to average ranks obtained on random, same-sized sets, the aver-age rank for GO:0019883 is extremely unusual. Compared to the statistic we would compute on GO:0019883 if viral expression is not associated with host expression, the observed average rank is modestly significant. A similar pattern recurs for many categories (Supplementary Figure 5). The two calibration approaches agree broadly but differ substantially in their ranking of categories, which suggests that the distinct enrichment signal is identified by the random-set approach.

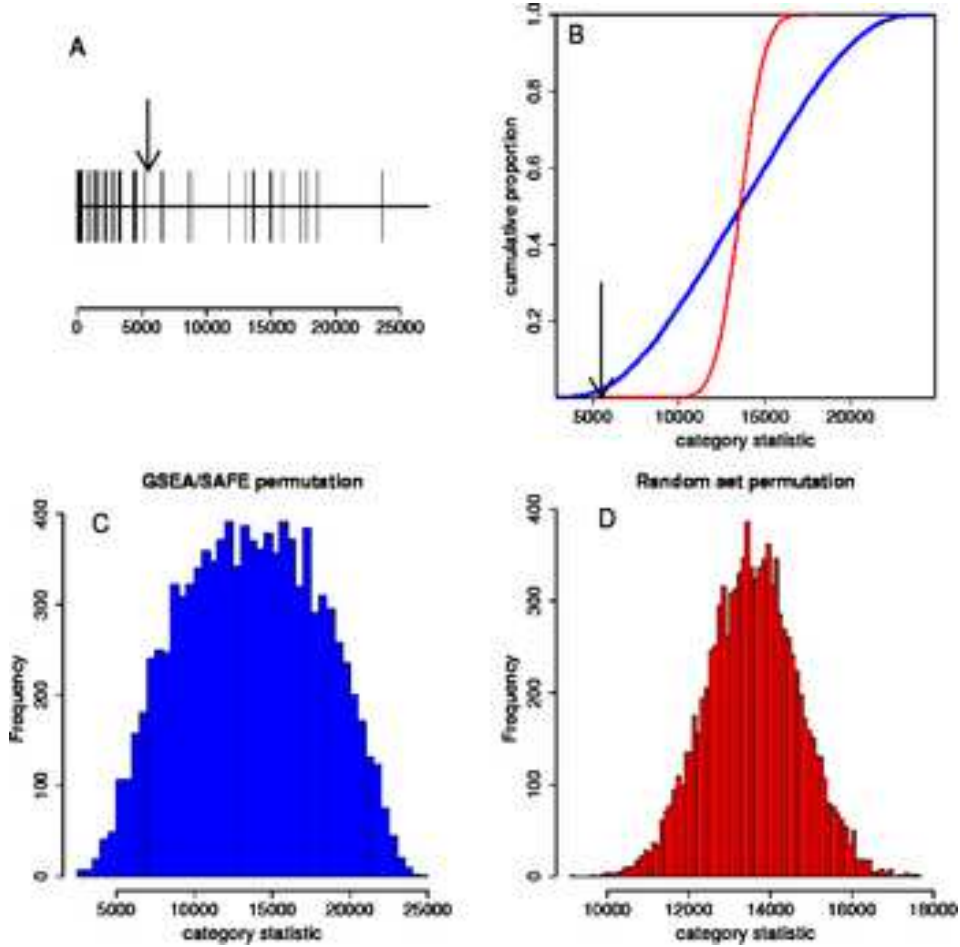

Fig. 3. *Comparison with SAFE/GSEA: Panel* A *is a rank plot of probe set correlation scores for the host-virus example. The positions of the* $m = 48$ *probe sets from GO:0019883 are marked, and suggest a preponderance of correlations on the negative side. The average rank is noted with an arrow (5478). Panel* B *compares the empirical distribution functions (e.d.f.s) of two different null distributions for the average rank statistic (red = random sets; blue = SAFE). Associated histograms are in panels* C *and* D*, but the e.d.f.s reveal the scale difference more dramatically;* $B = 10{,}000$ *random draws are used in each case. Recall that the random set (red) distribution is obtained by shuffling the GO:0019883 ranks in panel* A*. By contrast, the SAFE (blue) distribution returns to the original data and shuffles chip labels (as in Figure 3, Supplementary material). The p-value from SAFE is* $p = 0.02$ *and from random sets is* $p < 10^{-10}$ *(*$z = -7.1$*) based on a normal approximation to panel* D*.*

**4. Averaging or selection? A theoretical comparison.** In the preceding case study we observed empirical characteristics of two random-set methods for scoring category enrichment. The selection approach begins with a short list of extremely altered genes and asks if there is over-representation in the category. The averaging approach scores the category simply by averaging gene-level evidence across all genes in the category. The associated category $Z$ scores exhibit some positive correlation, but evidently they capture different components of the enrichment signal. Some theoretical findings are available which expose properties of the enrichment testing problem.

Our findings are developed in the context of a generic mixture model, one that is structurally similar to models commonly described in the microarray literature. The model is presented in order to develop a comparison of category scoring methods. It is not used for the analysis of data per se, but it sheds light on an interesting phenomenon created by this two-level (gene/category) inference problem. We find that each category-scoring method has its own domain of superiority in the state-space of enrichment problems; neither is always preferred. The result is somewhat surprising since information is obviously lost in the selection approach and not so obviously lost by averaging evidence. The result is related to the debate in statistics about model selection versus model averaging.

Consider genes $g \in \{1, 2, \ldots, G\}$ and quantitative gene-level scores $\{s_g\}$. The larger $s_g$, the more evidence for differential expression of gene $g$. A category $C$ is a known subset of $m < G$ genes sharing some particular biological function. The category may be scored for enrichment by one of two statistics:

$$\underbrace{\bar{X}_{\text{ave}} = \frac{1}{m} \sum_{g \in C} s_g}_{\text{averaging}}, \qquad \underbrace{\bar{X}_{\text{sel}} = \frac{1}{m} \sum_{g \in C} 1[s_g > k]}_{\text{selection}}.$$

To enable a comparison of the category scores, we frame the problem as a test of the null hypothesis that $C$ is not enriched. More specifically, suppose that each gene $g$ is either truly differentially expressed ($I_g = 1$) or not ($I_g = 0$) between the two cellular states. We allow that some fraction

$$\pi = \frac{1}{G} \sum_{g=1}^{G} I_g$$

of genes are truly differentially expressed. The category $C$ itself contains a fraction

$$\pi_C = \frac{1}{m} \sum_{g \in C} I_g$$

of differentially expressed genes. No enrichment means $H_0 : \pi_C = \pi$, and this is tested against the alternative $H_1 : \pi_C > \pi$. We can define enrichment simply as $\pi_C - \pi$. Lack of enrichment does not mean there is no differential expression; it just means there is not more than in the whole system. (One could also adjust for discreteness of $\pi_C$, but the adjustment would be negligible for modestly large category size $m$, and so it is not pursued.) Statistics $\bar{X}_{\text{sel}}$ and $\bar{X}_{\text{ave}}$ are two possible test statistics for testing $H_0$. We compare their power against various alternatives.

The latent differential expression indicator $I_g$ affects the distribution of the gene-level score $s_g$. A simple location model asserts that $s_g$ is normally distributed with unit variance and with mean $\delta I_g$ for a gene-level effect $\delta > 0$, and that all variables are independent. Normality is often reasonable for suitably transformed gene-level scores, such as $\log_2(\text{fold})$ or the transformed correlation (Section 3). Differential expression could potentially alter the variation of scores, but as a first approximation we focus on the location shifts only. The possible effects of among-gene dependence are important, but they are secondary in the present comparison of enrichment-scoring methods, hence, our demonstration is in the independence model (see discussion).

A test based on the average quantitative score $\bar{X}_{\text{ave}}$ uses the sampling distribution Normal$(\delta \pi_C, 1/m)$. Thus, the power of a level $\alpha$ test is $1 - \Phi(\tau_{\text{ave}})$, where $\Phi(\cdot)$ is the standard normal cumulative distribution and

$$\tau_{\text{ave}} = z_\alpha - \sqrt{m} \underbrace{(\pi_C - \pi)}_{\text{enrichment}} \underbrace{\delta}_{\text{effect}}. \tag{5}$$

Naturally, the power of the averaging approach increases with effect, enrichment and category size. One scenario is presented in Figure 4B: a category of size $m = 20$ is tested for enrichment at level $\alpha = 0.05$ in a system with $\pi = 0.2$ of genes differentially expressed.

The power of the selection approach is similarly derived. It entails a normal approximation for $\bar{X}_{\text{sel}}$ that is well justified in large categories by the central limit theorem. The power is $1 - \Phi(\tau_{\text{sel}})$, where

$$\tau_{\text{sel}} = z_\alpha \frac{\sigma(\pi)}{\sigma(\pi_C)} - \sqrt{m} \underbrace{(\pi_C - \pi)}_{\text{enrichment}} \underbrace{[\Phi(k) - \Phi(k - \delta)]/\sigma(\pi_C)}_{\text{effect}^*}. \tag{6}$$

Here $k = k(\pi, \delta, \tilde{\alpha})$ is chosen to deliver a FDR-controlled gene list at level $\tilde{\alpha}$ (see Appendix). Also, the variance function $\sigma^2(\pi_C)$ records the variance of $\sqrt{m}\bar{X}_{\mathrm{sel}}$. Figure 4A shows how the power to detect enrichment by the selection method is affected by $\pi_C - \pi$ and $\delta$ for categories of size $m = 20$, when $\pi = 0.2$ and $\alpha = \tilde{\alpha} = 0.05$.

The power surfaces in Figure 4 reveal an intriguing phenomena in enrichment testing. Both selection and averaging increase in power as either enrichment $\pi_C - \pi$ or effect $\delta$ increase. However, they increase differently, creating domains of superiority for each approach. The lower panels in Fig-ure 4 show these domains for the case indicated. When the $\pi_C - \pi$ is small, but $\delta$ is large, then it is better to use the selection approach. On the other hand, if $\delta$ is small, but $\pi_C - \pi$ is large, then it is better to average evidence across all genes in the category. The fact that selection can be superior is somewhat striking, since it entails a significant amount of information loss; each gene score is replaced by a binary indicator of whether or not the score is extremely large. On the other hand, if enrichment is weak, then averaging evidence combines a lot of noise with signal, thus, diminishing power.

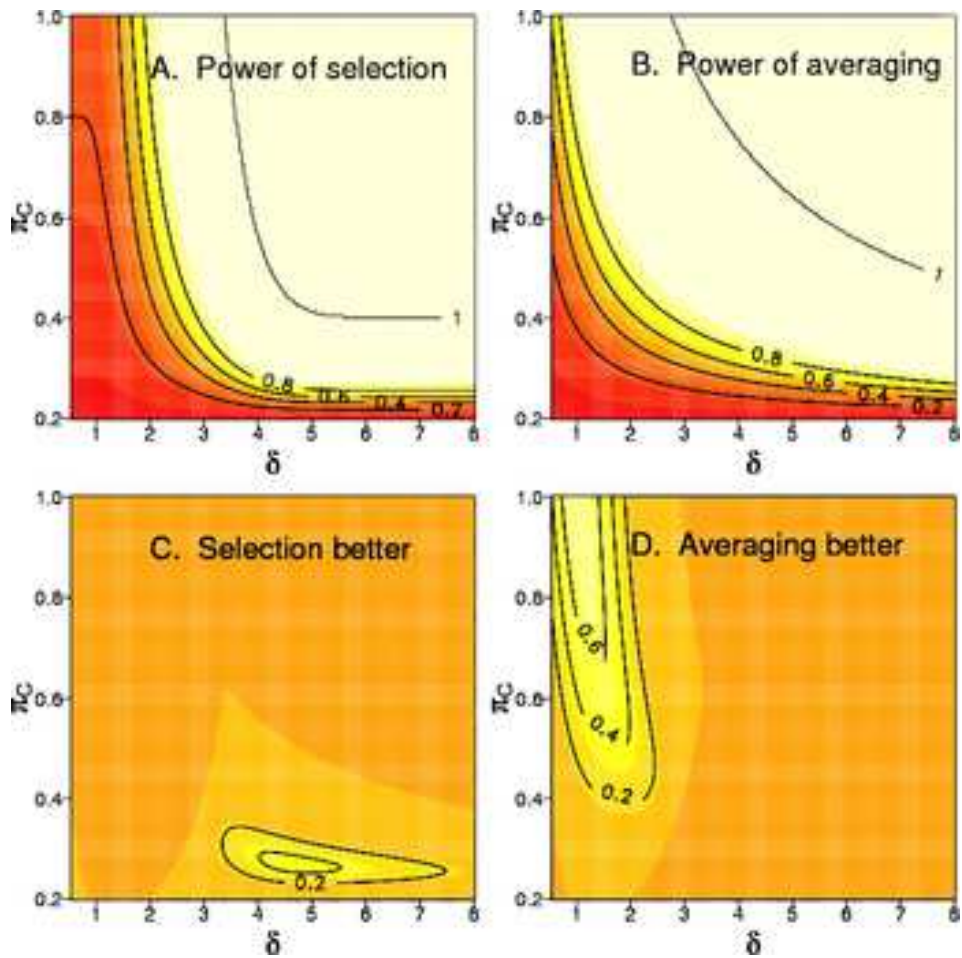

FIG. 4. *Power comparison. Power to detect enrichment is shown for a category of size* $m = 20$ *in a system with* $\pi = 0.2$ *differential expression overall (*A *and* B*). Selection and averaging are being compared, with low power in red and power increasing into the yellows and whites. The lower panels show domains of superiority for each method (by imaging the threshholded difference on a* $-1$*,* $1$ *scale). The enrichment is* $\pi_C - \pi$*, and the effect of differential expression per gene is* $\delta$*. The maximum power differential is 0.46 (*C*) and 0.74 (*D*).*

The nonlinearity of the power functions complicates a general comparison, but we have identified sufficient conditions for one or the other approach to be superior (see Appendix). To state the result, first put $\kappa = \tilde{\alpha}\pi/\{(1-\tilde{\alpha})(1-\pi)\}$, where, again, $\pi$ is the proportion of differentially expressed genes in the whole system and $\tilde{\alpha}$ is the FDR of the gene list used by the selection approach. We require $0 < \kappa < 1$, else it is not possible to have the desired FDR control; this is a weak condition, since it is implied if both $0 < \tilde{\alpha} < 1/2$ and $0 < \pi < 1/2$, for example.

THEOREM 1. *If $2\Phi^{-1}(\frac{1}{1+\kappa}) < \delta < \frac{1}{\sqrt{\kappa}} - \sqrt{\kappa}$, then for sufficiently large $m$, selection is more powerful than averaging. Also, given any $\pi_C > \pi$, there exists $\delta^*(\pi_C, \pi)$ such that if $0 < \delta < \delta^*(\pi_C, \pi)$, then for sufficiently large $m$, averaging is more powerful than selection.*

This is a finding about sub-optimality of two enrichment detection methods. In using a selection approach, there is limited power to detect enrichment when the category under test contains lots of genes that are altered by a small amount, reflecting the fact that these genes are not selected as the most significantly altered ones. By contrast, selection is superior to averaging if the category under test is enriched for a small number of highly altered genes. We note that the interval for $\delta$ in the first claim is nonempty when $\kappa$ is sufficiently small (say, smaller than 0.133). We also note that averaging is superior both when $\delta$ is very small and when $\delta$ is very large. Interesting power dynamics emerge in the interior of the state space, as, for instance, in Figure 4.

**5. Simultaneous inference with multiple categories.** An unsolved problem with enrichment calculations concerns the comparison of many categories that vary in size. On the null hypothesis of no enrichment, each $Z$ score is well calibrated by design, with zero mean and unit variance. But many categories may be enriched, and unlike simpler genomic testing problems, there is different power associated with these different tests. The distribution of $Z$ under the enrichment hypothesis is a function of both the unknown enrichment $\pi_C - \pi$ and the known category size $m$. For instance, in the location-shift model, $Z_{\text{ave}}$ has unit variance and mean $\sqrt{m}(\pi_C - \pi)\delta$. Owing to this size effect, the ranking of categories by $Z$ alone may not be optimal since large enriched categories will tend to have much larger $Z$ scores. The phenomenon is illustrated in the host-virus example in Figure 5A. The problem is not limited to $Z_{\text{ave}}$; it occurs too with $Z_{\text{sel}}$, especially when the selected set is relatively large (data not shown). A partial solution, demonstrated in Figure 5B, is to rank categories according to $Z/\sqrt{m}$; then categories are ranked by the estimated enrichment $\pi_C - \pi$. In itself this does not provide an error-controlled list of enriched categories; nor does it account for the nonconstant variance of $Z/\sqrt{m}$, but the ranking may be calibrated and remains useful for prioritizing categories across the GO networks.

A complete solution to simultaneous multiple-category testing will involve the joint distribution of category statistics, rather than their marginal distributions considered so far. Without developing calculations fully here, we note that the joint distribution of $Z$ scores across categories (conditional on $\{s_g\}$) is accessible by an analysis of intersections among the different categories. Proper-subset information, for example, is provided by the directed graphical structure of GO. Two elements of the random-set approach simplify the analysis of multiple categories. First, the permutation perspective described in Supplementary Table 1 carries over readily to multiple categories. The only difference is that we have an additional row for each category. The multiple-category information is equivalent to a (complicated) cross-classification of genes, and a permutation of the $\{s_g\}$, with the remaining tabulation fixed, is enough to generate the full joint distribution of category statistics (Supplementary Table 2). Second, valid and readily computed approximations to the joint distribution of category statistics are available. By restricting to moderate and large categories, $Z$ scores are approximately multivariate normal. The dependence is carried completely by between-category correlations, which can be computed by basic random-set methods [see Newton (2007)]. We find that if $Z_1$ and $Z_2$ are standardized category scores for two categories of sizes $m_1$ and $m_2$ which have an overlap of $m_{1,2}$ genes, then

$$\text{(7)} \qquad \text{corr}(Z_1, Z_2) = \frac{Gm_{1,2} - m_1 m_2}{\sqrt{m_1 m_2 (G - m_1)(G - m_2)}}.$$

For large $G$, this is approximately $m_{1,2}/\sqrt{m_1 m_2}$. In other words, dependence is induced by the overlap of categories, and increases the larger is this overlap.

A full analysis of the multiple-category testing problem is beyond the scope of this paper. However, we illustrate the utility of the correlation formula (7) and the multivariate normal approximation to describe one possible approach. Suppose there are $k$ categories under study. On the global null of no enrichment for any category, the vector $(Z_1, Z_2, \ldots, Z_k)$ of enrichment statistics is approximately Gaussian with mean zero, unit variances and covariances in (7). Ideally we would simulate the exact distribution by permuting gene scores as in the Supplementary Table 2, but the Gaussian provides a computational solution that is more convenient, especially with large $k$. By a Cholesky factorization of the known covariance matrix (even when it is less than full rank), we can easily simulate the multivariate normal vector. Sampled vectors respect the joint distribution of scores, for instance, as affected by category overlap, and provide input to various multiple testing schemes. For the NPC example, we report results of the $maxT$ procedure [Dudoit et al. (2003)] when the categories are ranked by $T = Z/\sqrt{m}$ to accommodate the effects of category size on power. For each simulated vector, we computed the maximum of the $T$ statistics across categories. Since we did not get $Z$'s by permutation, we did not need to recompute category statistics; rather we simply converted the $Z$'s to $T$'s using the category sizes. In the NPC example, 11 categories had a $T$ value exceeding the 95th percentile of the $maxT$ null distribution. A global view of the results is in Figure 5. Supplementary Table 3 summarizes the 11 interesting categories.

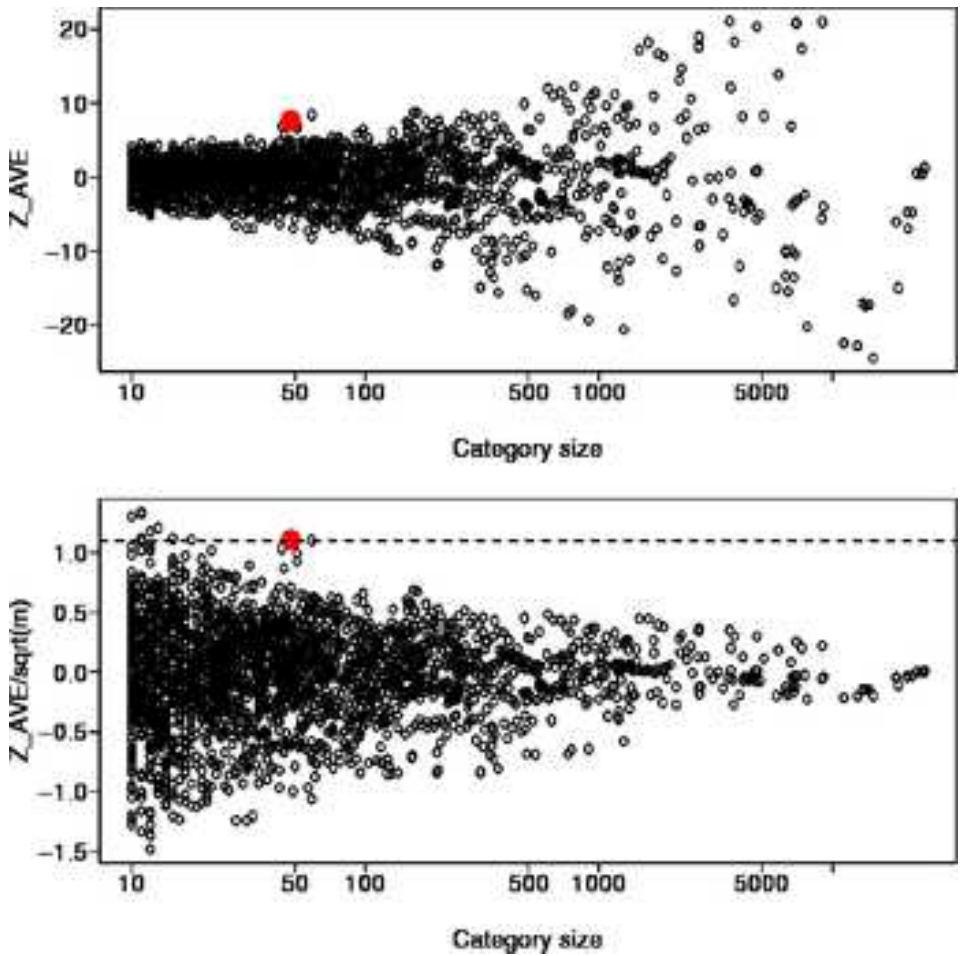

Fig. 5. *Power imbalance. Large enriched categories may get a strong Z score by virtue of their size m more so than by the level of enrichment $\pi_C - \pi$ (panel* A*). The simple correction $Z/\sqrt{m}$ provides a ranking of interesting categories that is adjusted for this size effect (panel* B*). Eleven categories, including GO:0019883 (red dot), have enrichment scores exceeding* 1.10 *(dashed line). These are significant at the* 5% *level by the* maxT *multiple testing correction computed from a multivariate normal simulation of category enrichment scores. Table 3 [Supplementary material] lists characteristics of these 11 interesting categories.*

**6. Discussion.** In analyzing functional categories related to nasopharyngeal cancer tissue, Sengupta et al. (2006) used the random-set enrichment method discussed here, with both binary selections from gene-level scores and averages of gene-level scores. The Supplementary Material associated with that paper presented the method; here, we have amplified that discussion, derived formulas (1) and (2) for standardization, evaluated the methodology empirically and theoretically, and provided comparative analyses. Evidence shows that the proposed category-scoring methods capture previously hidden components of the enrichment signal. Results are also provided that guide simultaneous inference across multiple categories.

The random-set calibration approach is the same one that underlies Fisher's exact test for independence between the selected gene list and the category under study. The simplicity of Fisher's test makes it compelling, but the test is limited by its focus on selected gene sets. Transferring random-set calculations to quantitative gene-level scores is complicated by the fact that the Fisher-test hypergeometric distribution no longer applies; an intractable distribution takes its place. In deriving the random-set mean and variance of a category score, we offer an easily computed approximation and standardized statistics for measuring category enrichment. This has several practical advantages over other schemes that use quantitative scores. By conditioning on results of the differential expression analysis, our calculations can handle a wide variety of output from that analysis and we need not revisit the raw data. Methods such as in Barry, Nobel and Wright (2005) and Subramanian et al. (2005) calibrate category scores by recomputing the differential expression results over random permutations of raw data. Not only can this be limited when the number of microarrays is small, but also the null hypothesis at work for such a permutation is the exchangeability of microarray labels, which asserts the absence of any differential expression. Insofar as enrichment concerns excess differential expression in a category rather than its absence, the random-set approach may be targeting enrichment more directly. Certainly the calibration is such that different aspects of the enrichment signal are being detected by the random-set approach. Further comparative analyses are warranted.

The issue of among-gene dependence is a subtle one that is relevant in enrichment calculations. The SAFE/GSEA permutation guards against ill-effects of such dependence by shuffling microarray labels and fixing whole profiles. Random-set scoring guards against these effects by conditioning on results of the differential expression analysis; since $\{s_g\}$ are fixed, whatever factors caused them to be dependent cannot enter the calculation. The flip side is that our interpretation of random-set $Z$ scores is focused on comparing the category statistic in hand to its hypothetical value from a random set (as opposed to its value on some hypothetical rerun of the whole expression study). Indeed, the dependence that in sampling theory terms would inflate the variance of the category score and speak against random sets is precisely the dependence that we aim to detect as the enrichment signal. Random-set calibration gains by its simple interpretation; still, one must take care that significant findings are attributable to biologically relevant enrichment rather than to something spurious.

In the context of a location-shift model for differential expression we compared two random-set methods in terms of their power to detect enrichment in a given category $C$. A clear picture of the sub-optimality of each method is available, pointing to the benefits of using several methods together to identify different aspects of the enrichment signal. When $C$ contains lots of modestly altered genes, then averaging evidence from all genes in the category is more powerful than selection; when $C$ is enriched just slightly, but the alteration effect is high, then the use of selected genes is preferred.

The random-set approach becomes feasible in multi-category inference because various properties can be explicitly calculated. Ranking categories by their $Z$ score normalized by the square root of category size gives a ranking based on estimated enrichment. This partially addresses the problem that large categories that are enriched at all will be detected with high probability, contrary, perhaps, to the aim to identify the most enriched categories. The joint distribution of $Z$ scores across multiple categories is induced by shuffling in a certain contingency table, and it is approximately multivariate normal under the complete null hypothesis. The dependence between scores is mediated by the extent of category overlap [equation (7)]. Multiple testing schemes can take advantage of this known dependence to assure the identification of significantly enriched categories, though optimal schemes remain to be worked out. We demonstrated the single-step $maxT$ approach in Section 5 and produced useful findings for the NPC study. Refinements are surely possible, either using stepwise approaches to control family-wise error or using one of the methods for false discovery rate control. It is tempting to convert the $Z$ scores to $p$-values and adopt one of these standard multiple testing adjustments, but size and dependence issues complicate a simple technology transfer. Further work in this direction may help to sort out reporting protocols when, for instance, multiple nodes in one branch of the GO graph exhibit varying degrees of enrichment.

## APPENDIX: SELECTION VS AVERAGING (THEOREM 1)

Which category-testing approach is more powerful depends on the sign of $\Delta = \tau_{\text{sel}} - \tau_{\text{ave}}$, the difference between the $Z$-score critical values corresponding to level $\alpha$ tests. Averaging is better if $\Delta > 0$; selection is better if $\Delta < 0$. Combining (5) and (6),

$$(8) \quad \Delta = z_\alpha \left\{ \frac{\sigma(\pi)}{\sigma(\pi_C)} - 1 \right\} - \sqrt{m}(\pi_C - \pi) \left\{ \frac{\mu_1 - \mu_0}{\sigma(\pi_C)} - \delta \right\},$$

where $\mu_0 = 1 - \Phi(k)$ is the mean of the Bernoulli trial $1[s_g > k]$ when $g$ is not differentially expressed, $\mu_1 = 1 - \Phi(k - \delta)$ is the mean when $g$ is differentially expressed, $\pi_C - \pi$ is the enrichment, $\delta$ is the effect of differential expression on $s_g$, $m$ is the category size, $\Phi(\cdot)$ is the standard normal cumulative distribution function, and $\sigma^2(\cdot)$ is a variance function. In this treatment gene scores $s_g$ are random, and the tests are marginal rather than conditional.

The threshhold $k = k(\pi, \delta, \tilde{\alpha})$ can be chosen to target a level $\tilde{\alpha}$ false discovery rate (FDR) owing to properties of the function

$$h(x) = \frac{1 - \Phi(x)}{1 - \Phi(x - \delta)}$$

for each fixed $\delta > 0$. By examining the derivative of $h(x)$ and invoking Mills' ratio [e.g., Gordon (1941)], one can prove the following:

LEMMA A.1. *$h(x)$ is monotone decreasing with $\lim_{x \to \infty} h(x) = 0$ and $\lim_{x \to -\infty} h(x) = 1$.*

Thus, $h(x)$ is invertible, and with fixed $\kappa = \tilde{\alpha}\pi/\{(1 - \tilde{\alpha})(1 - \pi)\} \in (0, 1)$, we can certainly find $k$ uniquely satisfying $h(k) = \kappa$. This $k$, furthermore, must define the threshhold for the level $\tilde{\alpha}$ FDR gene-selection procedure, since, by Bayes rule,

$$(9) \quad \tilde{\alpha} = P(H_0 | s_g > k) = \frac{\mu_0(1 - \pi)}{\mu_0(1 - \pi) + \mu_1 \pi},$$

where $\mu_0 = 1 - \Phi(k)$ and $\mu_1 = 1 - \Phi(k - \delta)$. Reorganizing (9), we get the defining relation $h(k) = \mu_0/\mu_1 = \kappa$. The 1–1 correspondence between $\mu_0$ (or $\mu_1$) and $\delta$ is further revealed through $\delta = \Phi^{-1}(1 - \mu_0) - \Phi^{-1}(1 - \mu_0/\kappa)$. Both $\mu_0$ and $\mu_1$ converge to 0 as $\delta \to 0$. As $\delta \to \infty$, $\mu_0 \to \kappa$ and $\mu_1 \to 1$. Always $\mu_0 = \kappa \mu_1$. Supplementary Figure 6 provides another view of these objects.

The behavior of the variance $\sigma^2(\pi_C) = \mu_0(1-\mu_0) + \pi_C\{\mu_1(1-\mu_1) - \mu_0(1-\mu_0)\}$ depends on the sign of the coefficient of $\pi_C$. Indeed, there is a critical point $\mu_0 = \kappa/(1+\kappa) < 1/2$, when this coefficient equals 0, and $\sigma^2(\pi_C) = \kappa/(1+\kappa)^2$ for all $\pi_C$. For $\mu_0 > \kappa/(1+\kappa)$, $\sigma^2(\pi_C)$ is decreasing in $\pi_C$; for $\mu_0 < \kappa/(1+\kappa)$, it is increasing in $\pi_C$. Noting the 1–1 correspondence of $\delta$ and $\mu_0$, we see the critical point occurs at $\delta = 2\Phi^{-1}(\frac{1}{1+\kappa})$.

Consider the first claim on superiority of selection. We seek conditions under which $\Delta < 0$. Figure 4 suggests looking at the case of large $\delta$. With $\delta > 2\Phi^{-1}(\frac{1}{1+\kappa})$, $\sigma^2(\pi_C)$ is decreasing in $\pi_C$, and so the first term in (8) is positive but bounded above by

$$z_\alpha\left(\frac{\sigma(\pi)}{\sigma(1)} - 1\right) > 0.$$

Being positive, this term works against having $\Delta < 0$, but the term is finite and not dependent on category size $m$. Thus, the second term in (8) dominates for sufficiently large $m$, and will drive $\Delta < 0$ as long as

$$\frac{\mu_1 - \mu_0}{\sigma(\pi_C)} > \delta. \tag{10}$$

With the effect $\delta$ larger than the critical point mentioned above, we have

$$\sigma^2(\pi_C) < \sigma^2(\pi) < \sigma^2(0) = \mu_0(1-\mu_0) < \kappa/(1+\kappa)^2$$

and so (10) holds if

$$\frac{\mu_1 - \mu_0}{\sqrt{\kappa/(1+\kappa)^2}} > \delta. \tag{11}$$

Noting $\mu_1 = \mu_0/\kappa$, rearranging (11) is equivalent to

$$\delta < \left(\frac{\kappa+1}{\sqrt{\kappa}}\right)\left(\frac{1-\kappa}{\kappa}\right)\mu_0,$$

and finally, noting $\mu_0 > \kappa/(1+\kappa)$ in this case, we obtain the upper bound $\delta < 1/\sqrt{\kappa} - \sqrt{\kappa}$. Thus, we have sufficient conditions establishing the superiority of selection over averaging in gene-set enrichment. Numerically we find the interval in Theorem 1 is nonempty for $0 < \kappa < 0.133$.

We prove the second claim similarly, and thus, in contrast to (10), we seek conditions under which

$$L = \frac{(\mu_1 - \mu_0)^2}{\sigma^2(\pi_C)} < \delta^2. \tag{12}$$

The left-hand side $L$ can be simplified, noting first that $\mu_1 = \mu_0/\kappa$ and also that $\sigma^2(\pi_C) = \mu_0\{1 + \pi_C(\frac{1}{\kappa} - 1)\} + O(\mu_0^2)$ as $\mu_0 \to 0$. Immediately we obtain

$$\lim_{\mu_0 \to 0} L/\mu_0 = \frac{(1/\kappa - 1)^2}{1 + \pi_C(1/\kappa - 1)} < \infty.$$

On the other hand, if the right-hand side of (12) diverges when normalized by $\mu_0$, then the stated inequality must hold for an interval of sufficiently small effects. Using Mills' ratio, we find the threshhold $k \approx (-\log \kappa)/\delta$ for small $\delta$ and, consequently,

$$\mu_0 = 1 - \Phi(k)$$
$$\approx \frac{\delta}{(-\log \kappa)\sqrt{2\pi}} \exp\left\{-\frac{1}{2}\frac{(\log \kappa)^2}{\delta^2}\right\}$$

for small $\delta$. Evaluating further shows that $\delta^2/\mu_0$ diverges as $\mu_0 \to 0$, completing the proof. $\square$

**Computing notes.** An R package, *allez*, was developed to implement random-set enrichment calculations, especially for GO categories. The source is available at http://www.stat.wisc.edu/~newton/. Calculations reported here used R [R Development Core Team (2005)] version 2.1 and Bioconductor [Gentleman et al. (2004)] package hug133plus2 version 1.10.0, built October 2005.

**Acknowledgments.** The authors gratefully acknowledge support from the National Cancer Institute (Grants CA64364, CA22443 and CA97944) and FONDECYT (Grants 1020712 and 1060729), and assistance provided by Christina Kendziorski (statistical analysis), Allan Hildesheim (NPC molecular study coordination), I.-How Chen (NPC tissue specimen collection), Chien-Jen Chen, Yu-Juen Cheng (NPC case-control study field effort), Robert Gentleman (category analysis), Deepayan Sarkar (R computing) and two anonymous referees for helpful comments. P. A. is a Howard Hughes Medical Institute Investigator.

M. A. Newton  
Departments of Statistics  
  and Biostatistics and Medical Informatics  
University of Wisconsin–Madison  
1300 University Avenue  
Madison, Wisconsin 53706  
USA  
E-mail: newton@stat.wisc.edu

F. A. Quintana  
Departamento de Estadística  
Facultad de Matemáticas  
Pontificia Universidad Católica de Chile  
Casilla 306, Correo 22  
Santiago  
Chile  
E-mail: quintana@mat.puc.ci

J. A. den Boon  
Institute for Molecular Virology  
and  
McArdle Laboratory for Cancer Research  
University of Wisconsin–Madison  
1525 Linden Drive  
Madison, Wisconsin 53706  
USA  
E-mail: jdenboon@wisc.edu

S. Sengupta  
WiCell Research Institute  
U.S. Mail P.O. Box 7365  
Madison, Wisconsin 53707-7365  
USA  
E-mail: ssengupta@wicell.org

P. AHLQUIST  
INSTITUTE FOR MOLECULAR VIROLOGY  
AND  
MCARDLE LABORATORY FOR CANCER RESEARCH  
AND  
HOWARD HUGHES MEDICAL INSTITUTE  
UNIVERSITY OF WISCONSIN–MADISON  
1525 LINDEN DRIVER  
MADISON, WISCONSIN 53706  
USA  
E-MAIL: ahlquist@wisc.edu